\documentclass[aps,prl,twocolumn,superscriptaddress,amsmath,amssymb]{revtex4-1}

\usepackage[colorlinks,pdfusetitle,urlcolor=blue,citecolor=blue,linkcolor=blue,bookmarksnumbered,plainpages=false]{hyperref}

\usepackage{color}
\usepackage{xcolor}

\usepackage{bm}
\usepackage{nicefrac}
\usepackage{siunitx}
\usepackage{soul}

\usepackage{wasysym}
\usepackage{graphicx}
\usepackage{braket}

\newcommand{\B}[1]{\mathbf{#1}}

\newcommand{\tens}[1]{\mathbb{#1}}

\newcommand{\mediator}{\mathbf{r}_\text{M}}

\newcommand{\donorFreq}{\mathbf{\omega}_\text{D}}

\begin{document}

\title{The virtual photon approximation for three-body interatomic Coulombic decay }

\author{Robert Bennett}
\affiliation{Physikalisches Institut, Albert-Ludwigs-Universit\"at Freiburg, Hermann-Herder-Str. 4, D-79104 Freiburg i. Br., Germany}
\affiliation{Freiburg Institute for Advanced Studies (FRIAS) Albertstr. 19, 79104 Freiburg, Germany}

\author{Petra Votavov\'{a}}
\affiliation{Charles University, Faculty of Mathematics and Physics, Institute of Theoretical Physics, V Hole\v{s}ovi\v{c}k\'ach 2, 180 00 Prague, Czech Republic}

\author{P\v{r}emysl Koloren\v{c}}
\affiliation{Charles University, Faculty of Mathematics and Physics, Institute of Theoretical Physics, V Hole\v{s}ovi\v{c}k\'ach 2, 180 00 Prague, Czech Republic}

\author{Tsveta~Miteva}
\author{Nicolas Sisourat}
\affiliation{Sorbonne Universit\'e, CNRS, Laboratoire de Chimie Physique - Matiè\`ere et Rayonnement, F-75005 Paris, France}

\author{Stefan Yoshi Buhmann}
\affiliation{Physikalisches Institut, Albert-Ludwigs-Universit\"at Freiburg, Hermann-Herder-Str. 4, D-79104 Freiburg i. Br., Germany}
\affiliation{Freiburg Institute for Advanced Studies (FRIAS) Albertstr. 19, 79104 Freiburg, Germany}

\date{\today}

\begin{abstract}

Interatomic Coulombic decay (ICD) is a mechanism which allows microscopic objects to rapidly exchange energy. When the two objects are distant, the energy transfer between the donor and acceptor species takes place via the exchange of a virtual photon. On the contrary, recent \textit{ab initio} calculations have revealed that the presence of a third passive species can significantly enhance the ICD rate at  short distances due to the effects of electronic wave function overlap and charge transfer states [Phys. Rev. Lett. 119, 083403 (2017)]. Here, we develop a virtual photon description of three-body ICD, showing that a mediator atom can have a significant influence at much larger distances. In this regime, this impact is due to the scattering of virtual photons off the mediator, allowing for simple analytical results and being manifest in a distinct geometry-dependence which includes interference effects. As a striking example, we show that in the retarded regime ICD can be substantially enhanced or suppressed depending on the position of the ICD-inactive object, even if the latter is far from both donor and acceptor species.

\end{abstract}

\maketitle

Interatomic Coulombic decay (ICD) is an ultrafast process by which energy can be transferred between microscopic objects (e.g. atoms, ions, clusters, quantum dots). First predicted just over two decades ago \cite{Cederbaum1997}, it involves an excited donor species which then decays and transmits sufficient energy to a neighbouring acceptor species that the latter can be ionised. Since most of the excess energy of the donor is spent ejecting an electron from the acceptor, a slow electron is left in the continuum \cite{Gokhberg2014}. As well as being one of the experimental signatures of ICD \cite{Marburger2003}, it has been shown that such slow electrons can be damaging in a biological context \cite{Boudaffa2000}.

The ICD rate is an important property in characterisation of the process. However, its computation is a challenging task. Most calculations of ICD rates use techniques adapted from computational quantum chemistry, necessitated by the donor and acceptor species being very closely spaced so that orbital overlap has a dramatic effect on the system \cite{Averbukh2004,Averbukh2005}. However, at slightly larger distances it is possible to use a `virtual photon approximation' \cite{Averbukh2004}. There, the donor and acceptor are considered as separate objects coupled via the quantised electromagnetic field. This results in a simple analytic expression for the rate that depends on the single-body decay rate of the donor, the photoionisation cross-section of the acceptor and their mutual separation. This expression is often used as a consistency check for the large-distance behaviour of a particular quantum chemical calculation. Furthermore, an analytical formula for the ICD rate provides a simple means to investigate large systems based on the decomposition of the clusters into pairs~\cite{Fasshauer2014a,Fasshauer2016}.

Recently, a type of three-body ICD mechanism known as superexchange was proposed \cite{Miteva2017}. Based on extensive \textit{ab initio} calculations, it was shown that the rate of energy transfer can be substantially enhanced in the presence of a third ICD-inactive mediating atom. However, there is no equivalent of the virtual photon approximation for the three-body ICD process.  This problem can be solved by making use of a macroscopic quantum electrodynamics (QED) based approach recently put forward in \cite{Hemmerich2018} where the effects of the environment near the decaying pair can be accounted for. It should be mentioned that the corresponding situation for F\"{o}rster resonant energy transfer (FRET) has been investigated previously \cite{Daniels2002,Salam2012}.

In this Letter we develop the virtual photon approximation for three-body ICD and find agreement with ab initio data in the relevant regimes. The new theory allows us to readily investigate retardation and geometrical effects in three-body ICD, providing insight into long-range energy transfer processes and guiding future \textit{ab initio} calculations. Our method is based on the recently-derived formula for the ICD rate in a generic medium \cite{Hemmerich2018};
\begin{equation}\label{ICDFormula}
\Gamma =  2\pi^2\!\!\!\!\sum_{\text{channels}} \!\!\!\!\gamma_\text{D}\sigma_\text{A}(\hbar \donorFreq)\text{Tr} [\mathbb{G}(\mathbf{r}_\text{A},\mathbf{r}_\text{D},\omega_\text{D})  \cdot \mathbb{G}^* (\mathbf{r}_\text{D}, \mathbf{r}_\text{A},\omega_\text{D})],
\end{equation}
where $\gamma_\text{D}$ is the free-space decay rate of the donor species and $\sigma_\text{A}(\hbar \donorFreq)$ is the photoionisation cross--section of the acceptor. The quantity $\tens{G}(\mathbf{r},\mathbf{r}',\omega) $ is the Green's tensor of the Helmholtz equation, describing propagation of excitations of frequency $\omega$ from point $\B{r}'$ to $\B{r}$, taking into account the effects of any environment that may exist between or around the donor and acceptor. 

\begin{figure}
\centerline{\includegraphics[width=\columnwidth]{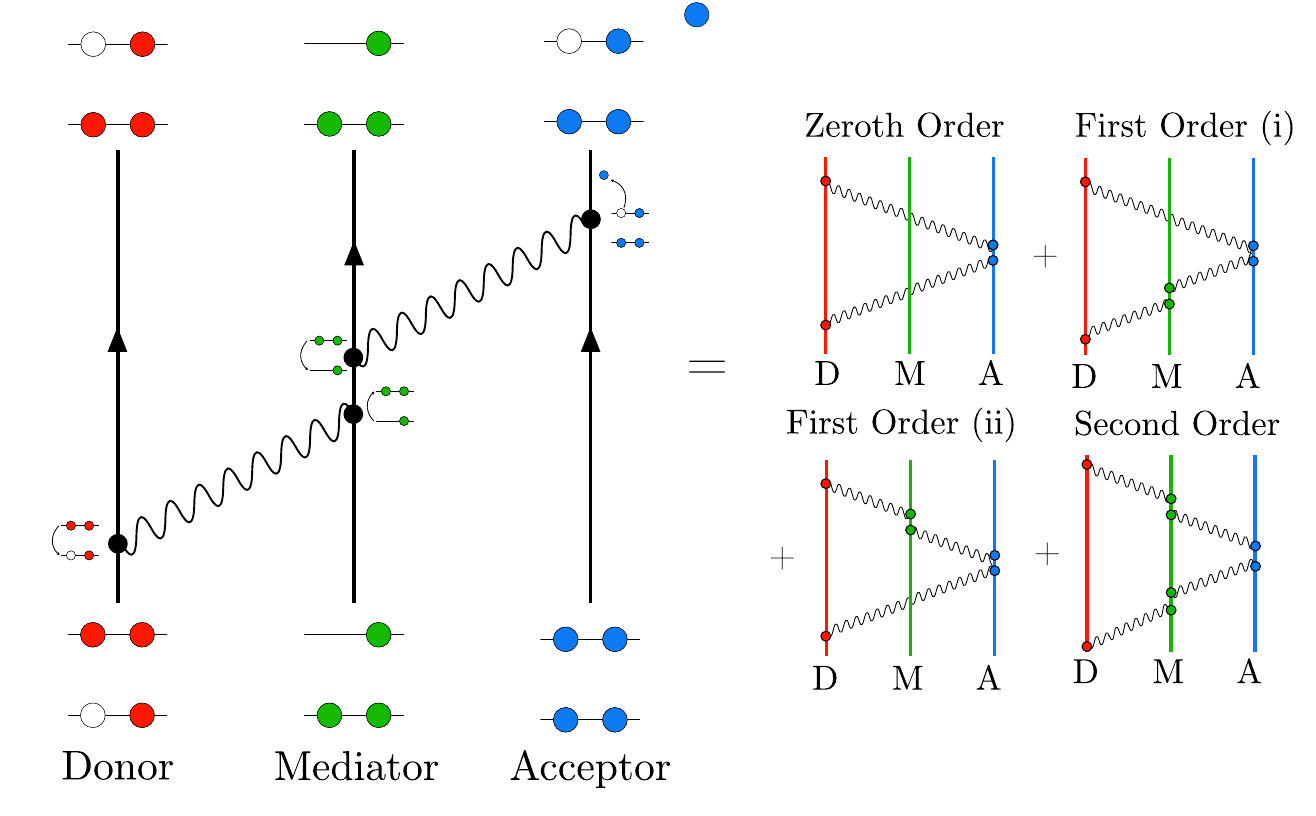}}
\caption{The process we consider. An excited donor relaxes to its ground state, emitting a photon which eventually ionises an acceptor. On the way, this photon may interact with a mediating atom, whose contribution is considered perturbatively as far as is consistent with the perturbation theory that leads to Eq.~\eqref{ICDFormula}  }
\label{ProcessIllustration}
\end{figure}

We consider a process where an ICD-inactive atom absorbs the virtual photon emitted from the donor and then re-emits this photon, which is subsequently absorbed by the acceptor, ejecting an electron and finishing the ICD process, as illustrated in Fig.~\ref{ProcessIllustration}.

\begin{figure}
      \includegraphics[width=\columnwidth]{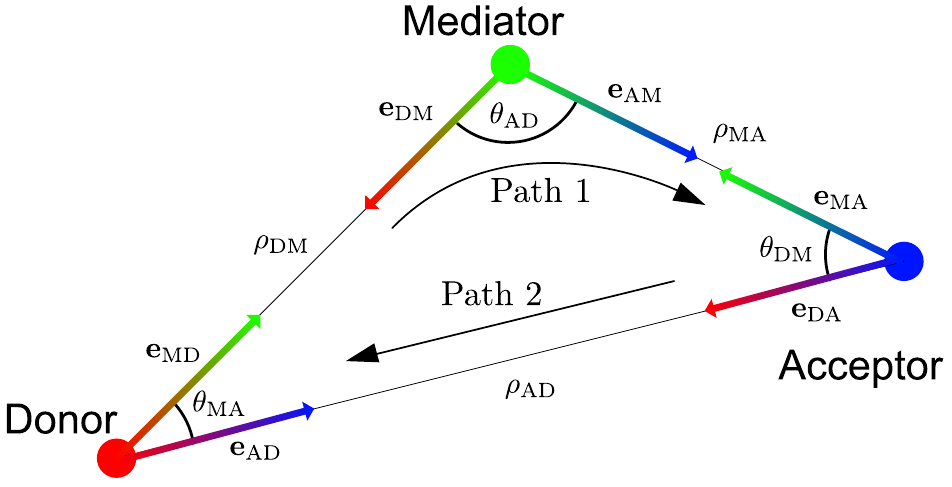}
        \caption{Definitions of geometrical quantities. Path 1 and Path 2 indicate the forward and backward parts of the closed trajectory corresponding to the first order diagram (i) (cf. Fig. \ref{ProcessIllustration})}\label{Triangles}
    \end{figure}

In order to calculate the rate in this situation, we need an explicit form of the Green's tensor for an environment consisting of a single atom. To obtain this we expand the Green's tensor in a Born series around a known `background' Green's tensor $\tens{G}^{(0)}$ \cite{Buhmann2006}, which could represent vacuum, a homogeneous medium, a dielectric plate or any other geometry for which the Green's tensor is analytically known. We have:
\begin{align} \label{BornSeriesFirstOrder}
& \tens{G}\left( \mathbf{r},\mathbf{r}',\omega  \right)=  \tens{G}^{(0)}\left( \mathbf{r},\mathbf{r}',\omega  \right)\notag \\
& + \mu _0 \omega ^2\int \text{d}^3 \B{s}\, {n \left( \B{s} \right) \tens{G}^{(0)}\left( \mathbf{r},\B{s},\omega  \right) \cdot \bm{\alpha }\left( \omega  \right)  \cdot \tens{G}\left( \B{s},\mathbf{r}',\omega  \right)}
\end{align}
where $\bm{\alpha }(\omega)$ is the polarisability tensor of the mediating atom at frequency $\omega$, and $\mu_0$ is the vacuum permeability. The integral runs over the volume of any dielectric bodies not included in $\tens{G}^{(0)}$, each of which has a position-dependent atomic number density $n(\B{r})$. Eq.~\eqref{BornSeriesFirstOrder} is exact but infinitely recursive, nevertheless a result for any order can be found by repeated substitution. For example the first order approximation can be obtained by substituting $\tens{G} \to \tens{G}^{(0)}$ on the right hand side of \eqref{BornSeriesFirstOrder}. The result is the Green's tensor for the background environment described by $\tens{G}^{(0)}$, with a single atom added. The terms in this approximation can be visualised as scattering from $\B{r}'$ to $\B{r}$ via intermediate scattering points $\B{s}$ with coupling strength determined by $\bm{\alpha }(\omega)$. In our system the mediator is a single atom in vacuum at position $\mediator$, which we describe via a Dirac delta function number density $n(\B{s}) = \delta (\B{s}-\mediator)$. Using this in the first-order approximation to the Born series \eqref{BornSeriesFirstOrder} we have;
\begin{align} \label{GApprox}
 \tens{G}^{(1)}&\left( \mathbf{r},\mathbf{r}',\omega  \right)=  \tens{G}^{(0)}\left( \mathbf{r},\mathbf{r}',\omega  \right)\notag \\
& + \mu _0 \omega ^2\tens{G}^{(0)}\left( \mathbf{r},\mediator,\omega  \right) \cdot \bm{\alpha }\left( \omega  \right)  \cdot \tens{G}^{(0)}\left( \mediator,\B{r}',\omega  \right)
\end{align}
The higher-order terms depend on self-interactions, corresponding to quantities like $\tens{G}^{(0)}\left( \mediator,\mediator,\omega  \right)$. These are already taken into account by using an observed polarisability that includes QED corrections \cite{Schmidt2007}, meaning that Eq.~\eqref{GApprox} is in principle an exact relation. Crucially, $\tens{G}^{(1)}$ now only depends on the vacuum Green's tensor $\tens{G}^{(0)}$ and the polarisability $\bm{\alpha}$, which are both well-known. Substituting the Green's tensor \eqref{GApprox} into the rate formula \eqref{ICDFormula}, one finds three types of term which are of zeroth, first and second order in the polarisability. One can then proceed to use the vacuum Green's tensor in these four terms and work out ICD rates for arbitrary arrangements of donor, mediator and acceptor. However, the resulting expressions are extremely complex and lengthy, so do not provide much insight or intuition. We can considerably simplify calculations by anticipating that the transition wavelength of the donor we consider is far longer than the few angstroms at which ICD processes are active. This means we are in the non-retarded (static) regime, in which the Green's tensor is given by (see, for example \cite{Buhmann2012a});
 \begin{equation}\label{VacGNR}
\tens{G}^{(0)}_\text{NR}\left( \mathbf{r},\mathbf{r}',\omega  \right)  = -\frac{c^2}{4\pi\omega^2 \rho^3} \left(\mathbb{I} - 3 \B{e}_\rho \otimes \B{e}_\rho\right)
\end{equation}
where $\rho = |\B{r}-\B{r}'|$, and $ \B{e}_\rho $ is a unit vector in the direction of $\B{r}-\B{r}'$.
We also simplify the derivation by assuming that the mediator has a real, isotropic and frequency-independent polarisability $\bm{\alpha }(\omega)=\alpha \mathbb{I}$. It is also useful to work with the polarisability volume $\alpha/(4\pi \varepsilon_0)$ (where $\varepsilon_0$ is the vacuum permittivity) rather than the polarisability itself so for the rest of this article we make the replacement $\alpha/(4\pi \varepsilon_0)\mapsto \alpha$. Substituting the non-retarded Green's tensor \eqref{GApprox} into the rate formula \eqref{ICDFormula}, we find;
\begin{align}\label{TotalRateNonRet}
& \Gamma_\text{NR}=C_6  \Bigg[\frac{1}{\rho_\text{AD}^6}+\frac{3\alpha^2}{2\rho_\text{MD}^6\rho_\text{MA}^6}\left(1+ \cos^2 \theta_{\text{AD}}\right)\notag \\
 &- \frac{ \alpha}{\rho^3_\text{AD}\rho^3_\text{DM}\rho^3_\text{MA}}\left(1+3 \cos\theta_\text{DM}\cos\theta_\text{MA}\cos\theta_\text{AD}\right)\Bigg]
\end{align}
where we have defined a Hamaker-type coefficient
$C_6 = \gamma_\text{D}\sigma_\text{A}(\hbar \donorFreq)  \frac{ 3c^4}{ 4\donorFreq^4}$
and written the result in terms of the angles and distances defined in Fig.~\ref{Triangles}. 

An important special case can be extracted from the general non-retarded rate \eqref{TotalRateNonRet}. This is the rate $\Gamma^\text{L}_\text{NR}$ for a colinear arrangement, in which $\theta_\text{MA}=\theta_\text{DM}=0$ and $\theta_\text{AD}=\pi$. Then;
\begin{align}\label{TotalRateNonRetColinear}
 \Gamma^\text{L}_\text{NR}&=\frac{C_6}{\rho_\text{AD}^6}\left(1+\frac{2}{3}u_\text{NR}+u_\text{NR}^2\right)
\end{align}
where 
$ u_\text{NR}= \alpha\rho^3_\text{AD}/({\rho^3_\text{DM}\rho^3_\text{MA}})$
is a dimensionless number indicating the strength of the interaction with the mediator, which must be less than unity for our perturbative approach to be applicable. For comparison with recent \textit{ab initio} work \cite{Miteva2017}, we further assume that the mediator is halfway between the donor and acceptor ($\rho_\text{DM}=\rho_\text{MA}=\rho_\text{AD}/2$), giving a very simple result;
\begin{align}\label{TotalRateNonRetColinearHalfway}
 \Gamma^\text{mid}_\text{NR}&=C_6  \left(\frac{1}{\rho_\text{AD}^6}+ \frac{128 \alpha}{\rho_\text{AD}^9}+\frac{12 \, 288\,  \alpha^2}{\rho_\text{AD}^{12}}\right). 
\end{align}

Formula \eqref{TotalRateNonRetColinearHalfway} can now be compared to \textit{ab initio} calculations. As in \cite{Miteva2017}, we consider the case where the donor and acceptor are both neon, and the mediator is helium. Before making this comparison, however, we note that in \cite{Miteva2017} excited configurations of the type Ne$^+$He$^*$Ne were excluded from the calculations, meaning we should consider only the static polarisability of the helium, given by $\alpha_\text{He}=0.205 \text{\AA}^3$\cite{Rohrmann2018}.

We also need a value for the two-body coefficient $C_6$, which can in principle be calculated from known values of the free-space decay rate of the donor $\gamma_\text{D}$, the photoionization cross section of the acceptor $\sigma_\text{A}(\hbar \donorFreq)$ and the transition frequencies involved in the process $\omega_\text{D}$. Indeed this can be done for the system of interest here with results coinciding up to small numerical factors, but due to complications of the type discussed in \cite{Averbukh2004}, we determine the $C_6$ from \textit{ab initio} calculations. We do this by removing the mediator from the system, and place the neon atoms far enough apart that a $1/\rho_\text{AD}^6$ distance-dependence is seen. 

\begin{figure}[t!]
      \includegraphics[width=\columnwidth]{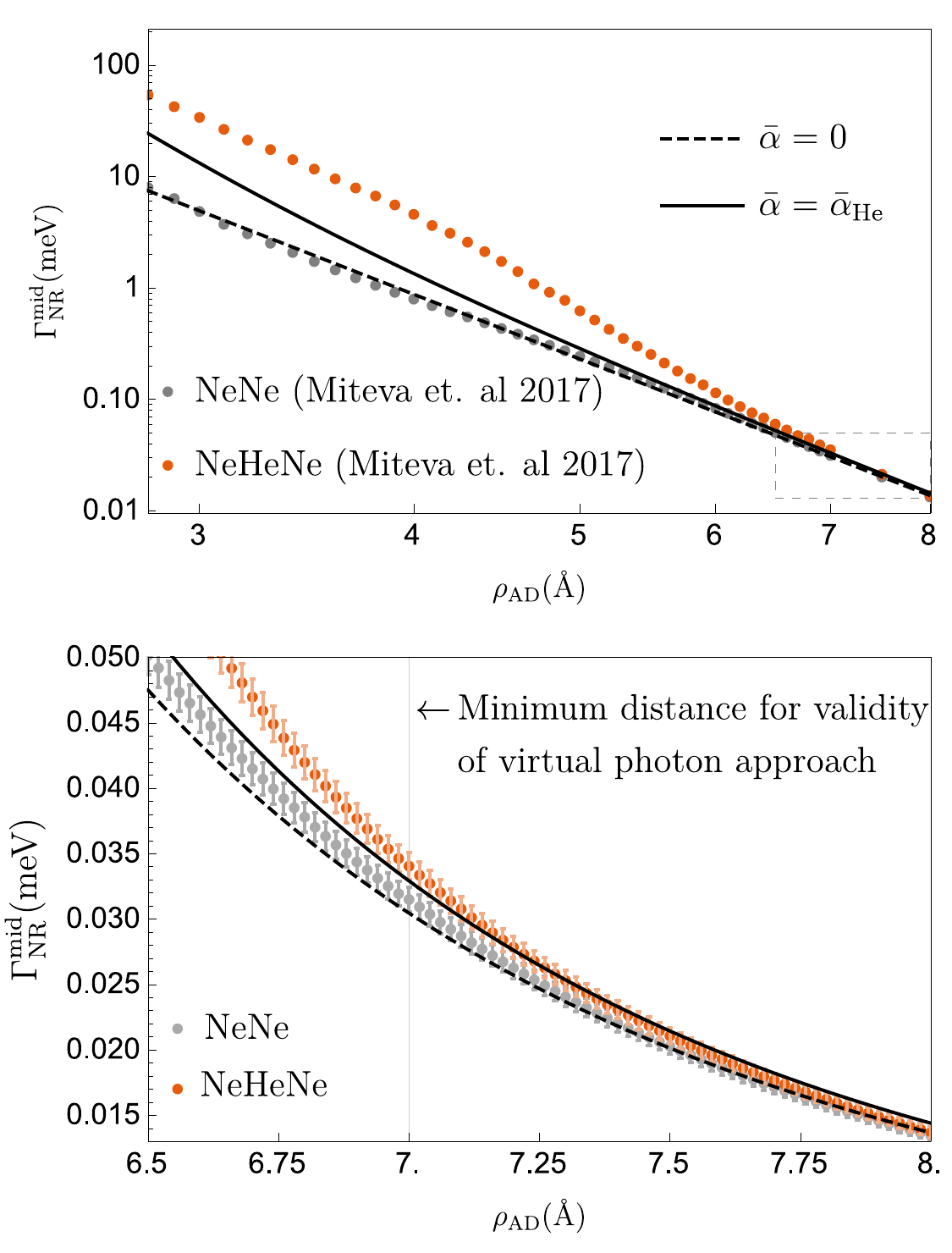}
        \caption{ICD rate vs donor-acceptor distance, with the mediator placed at the midpoint, $C_6=3.6$eV and $\alpha=\alpha_\text{He} = 0.205 \text{\AA}^3$. NeNe denotes the situation when the mediator is removed entirely, NeHeNe in the upper plot represents the same data as presented in \cite{Miteva2017}, while the lower plot contains new high-resolution data calculated for this work. The error bars on the ab initio data are 3\%, which comes from the standard deviation of the decay widths calculated by Fano-ADC-Stieltjes method.}\label{CombinedPlot}
    \end{figure}

Shown in Fig.~\ref{CombinedPlot} is the comparison between the ICD widths given in our new approach by  Eq.~\eqref{TotalRateNonRetColinearHalfway}, and the calculated with the \textit{ab initio} Fano-algebraic diagrammatic construction (ADC)-Stieltjes method~\cite{Averbukh2005,Kolorenc2015} (see \cite{Miteva2017} for details of the calculations). As seen in the upper panel, the results deviate significantly from the \textit{ab initio} data if $\rho_\text{AD}\lesssim 7$\AA. This is to be expected for at least two reasons. Firstly, the virtual photon method should fail when there is significant wave function overlap, as discussed in detail in \cite{Averbukh2004}. Secondly, the superexchange enhancement seen below 7\AA\,  in \cite{Miteva2017} relies on intermediate states that include charge transfer, where the helium gains an electron to become He$^-$, which are not included in our virtual photon approach. The contribution of these charge transfer intermediate states to three-body ICD decreases exponentially with the neon-neon distance, we therefore concentrate on distances larger than $7$\AA. As shown in the lower panel of Fig.~\ref{CombinedPlot}, the ICD widths obtained with both approaches agree well, supporting the general approach taken here.
It should be mentioned that even without the inclusion of the charge transfer intermediate states a clear enhancement of the ICD rates is seen. In our approach, any mediator-dependence of the rate is to be understood as coming from the mediator's modification of the electromagnetic field that couples the donor and acceptor species, rather than modification of atomic properties themselves.

\begin{figure}[h!]
      \includegraphics[width=\columnwidth]{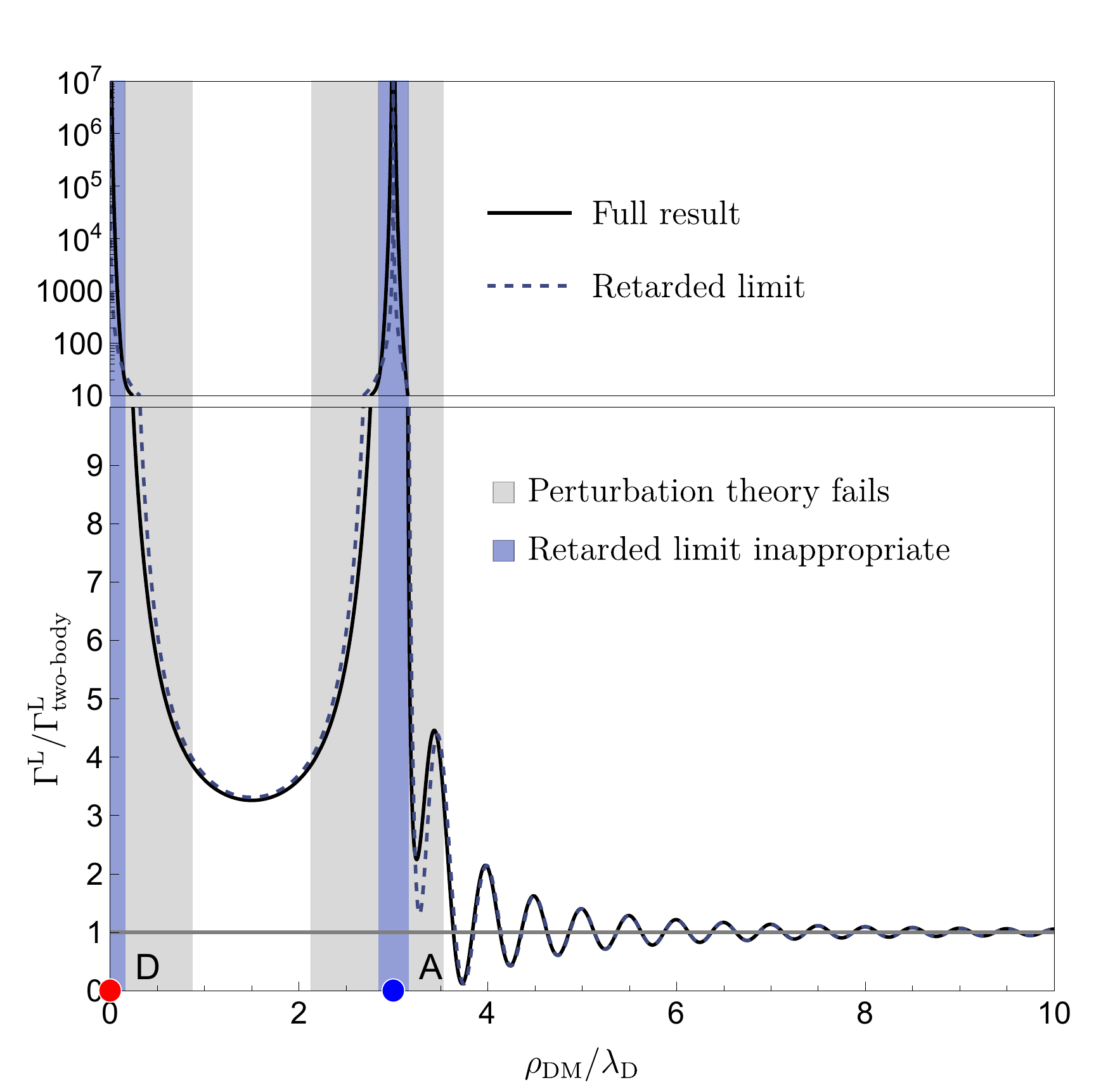}
            \includegraphics[width=\columnwidth]{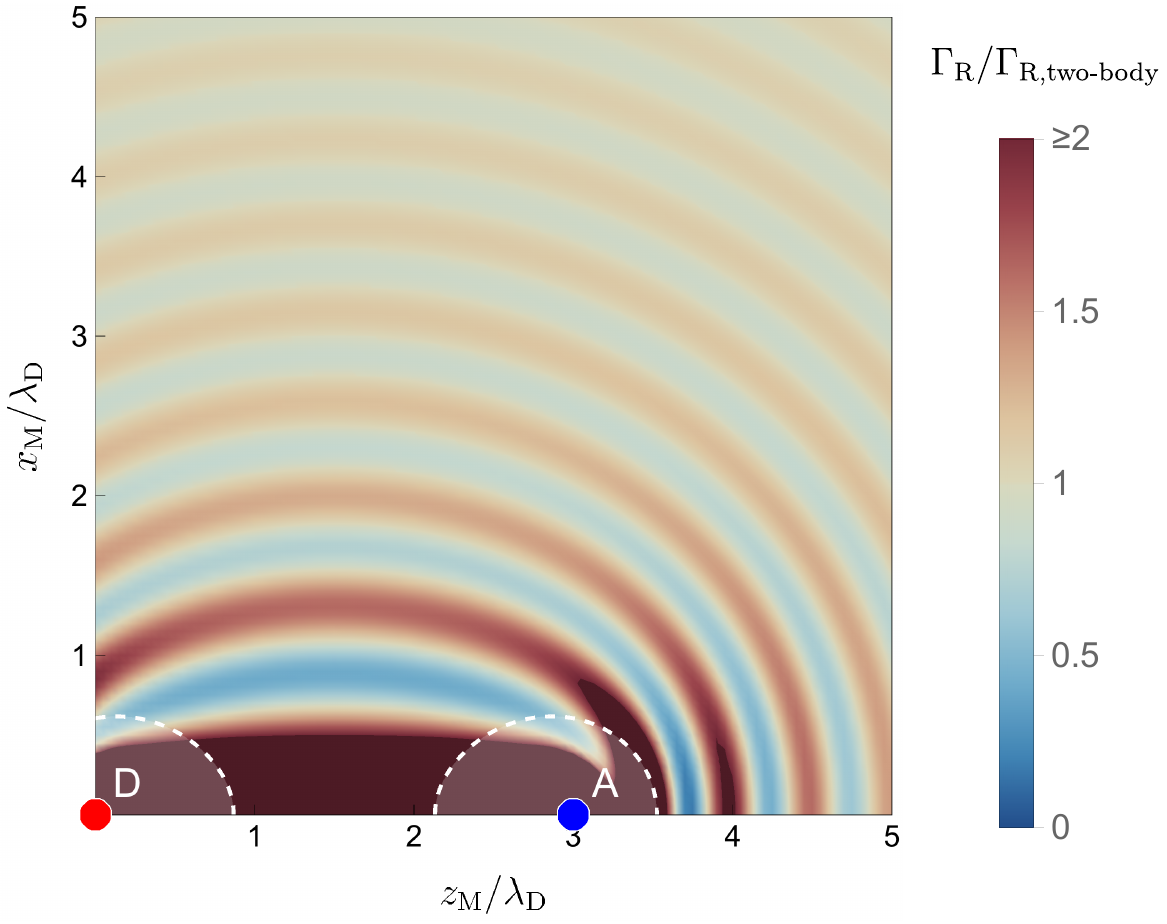}
        \caption{(Upper) Three-body far-field ICD rate $\Gamma^\text{L}_\text{R}$ given by Eq.~\eqref{GammaR} (dashed lines), normalised to the two-body rate given by setting $\alpha$ to zero in that equation. The donor is at the origin, the acceptor is placed three wavelengths $\lambda_\text{D}$ away ($\rho_\text{AD}=3\lambda_\text{D}$), and the polarisability volume is $(\lambda_\text{D}/4)^3$. Also shown as a solid line is the full result at any distance (neither retarded or non-retarded), showing that the retarded limit \eqref{GammaR} is valid further than $\lambda_\text{D}/(2\pi)$ from donor or acceptor. However, the perturbative approach used here becomes unreliable if $u_\text{R}>1$, which turns out to be a more stringent condition than the retarded limit, as shown.   Shown in the lower panel is the plot of the generalisation of Eq.~\eqref{GammaR} to two dimensions using the same parameters and placing the acceptor on the $z$ axis. There the regions bounded by white dashed lines represent the region where our perturbation theory becomes unreliable.}\label{Plot1D}
    \end{figure}

All the results shown so far are in the non-retarded regime, as can be seen by noting that photon frequencies in \cite{Miteva2017} are determined by the 2s$^{-1} \to$ 2p$^{-1}$ transition of Ne$^+$, which has a wavelength of 480\AA. Retardation sets in at the transition wavelength divided by $2\pi$, which for this system is an order of magnitude longer than all considered separations of donor, mediator and acceptor considered so far. Nevertheless, since the method used here intrinsically includes retardation \cite{Hemmerich2018}, we can put the three-body ICD process  into a broader context by considering the consequences of using a donor with a higher transition frequency, or, equivalently, large spacing between the three atoms. Physical systems which may fulfil these criteria include highly-charged or hollow ions, as discussed in detail in our previous work \cite{Hemmerich2018}, and experimentally observed in the context of FRET in \cite{Ceolin2017}. One of the most striking consequences of retardation is that the ICD rate can oscillate in space if the mediator is placed anywhere other than on the line joining donor and acceptor. To see this, we use the Green's tensor in its retarded (far-field) limit as found in, for example, \cite{Buhmann2012a};

     \begin{equation}
\tens{G}^{(0)}_\text{R}\left( \mathbf{r},\mathbf{r}',\omega  \right)  =\frac{e^{i\omega \rho/c}}{4\pi \rho} \left(\mathbb{I} -  \B{e}_\rho \otimes \B{e}_\rho\right)
\end{equation}
in \eqref{ICDFormula} via \eqref{GApprox}, giving a rate $\Gamma_\text{R}$. For general mediator position this is a very long and unwieldy expression, so we only report an explicit formula for the rate $\Gamma^\text{L}_\text{R}$ in the colinear arrangement;
\begin{equation}\label{GammaR}
\Gamma^\text{L}_\text{R} =\frac{C_2}{\rho_\text{AD}^2} \!\left[ 1+u_\text{R}^2+2u_\text{R} \begin{cases}\cos \left({2 \omega_\text{D} \rho_\text{AM}}/{c} \right) &\!\text{if } \theta_\text{AD}=0\\
1 &\!\text{if } \theta_\text{AD}=\pi \end{cases} \right]
\end{equation}
where $u_\text{R}={{\alpha}\rho^2_\text{AD}  \donorFreq^2 }/({\rho_\text{AD}  \rho_\text{AM} \rho_\text{DM}}c^2)$ is a dimensionless number describing the strength of the interaction, $\theta_\text{AD}$ is defined in Fig.~\ref{Triangles} and $C_2=\gamma_\text{D}\sigma_\text{A}(\hbar \donorFreq)/4$. When the mediator is outside the region between donor and acceptor, spatial oscillations occur. This is shown in Fig.~\ref{Plot1D},
where we have plotted the rate \eqref{GammaR} for fixed donor and acceptor positions, and a mediator whose position is allowed to vary. It is remarkable that, in the retarded (far-field) regime, the ICD rate between donor and acceptor can be strongly suppressed by placing an ICD-inactive atom \emph{outside} the region between them. Such suppression comes from processes where the mediator interacts once with the electromagnetic field (i.e. the oscillatory term in \eqref{GammaR} is linear in $\alpha$). This, coupled with the fact that no oscillations occur if $\theta_\text{AD}=\pi$, demonstrates that the oscillations have their origin in the phase accumulated along the example trajectory indicated in Fig. \ref{Triangles}, which is determined by the phase difference between the (mediated) forward path 1 and the (direct) backward path 2.

In this Letter we have presented a virtual photon approximation for three-body ICD by taking advantage of the recently-introduced theoretical approach for calculation of the rate in arbitrary environments \cite{Hemmerich2018}. Our approach gives simple analytic results which are shown to fit well to independent \textit{ab initio} calculations using the Fano-ADC-Stieltjes method. Our approach also allows us to make predictions and give guiding principles concerning three-body ICD in other systems. Furthermore, we show that the ICD lifetime as a function of the distance can change dramatically depending on the relative values of the field-mediator coupling and the donor transition wavelength.

\acknowledgments{R.B. and S.Y.B. thank Akbar Salam for fruitful discussions. R.B. acknowledges financial support by the Alexander von Humboldt foundation and S.Y.B. thanks the Deutsche Forschungsgemeinschaft (grant BU 1803/3-1476). R.B. and S.Y.B. both acknowledge support from the Freiburg Institute for Advanced Studies (FRIAS). P.K. and P.V. acknowledges financial support by the Czech Science Foundation (Project GA\v{C}R No.17-10866S). This project has received funding from Agence Nationale de la Recherche through the program ANR-16-CE29-0016-01.} \\

\end{document}